# Physics Letters B publications from 1967 to 2020.
An analysis of the WEB page content of PLB

*by*
*Hendrik Weerts*
*Dept. of Physics & Astronomy Michigan State University and*
*HEP Division, Argonne National Laboratory*
*July 2020*

**Abstract**
Having been an editor for Physics Letters B (PLB) for many years I became interested in the history of publishing, especially in particle physics (HEP). Since PLB goes back to the 1960's and an index of all PLB publications is available online, this information was used to look at the history of PLB publications over time. It should be noted that PLB publishes new results in particle physics/high energy physics, nuclear physics and also in astrophysics & cosmology. This is the start of an effort to look at all publishers that publish research from HEP.

*Introduction*
The publisher Elsevier started publishing Physics Letters B(PLB) [1] to cover particle physics (HEP) and Nuclear Physics(NP) in 1967 with Volume 24 and Issue 1. Before that the journal was known as Physics Letters, covered basically all physics, was published from 1962 to 1966 and covered Volumes 1 through 23. After that it split into Physics Letters A and B. More information can be found in Wikipedia [2]. This note covers volumes 24 (first PLB volume) through 799 (last volume of 2019), covering 53 years of publications in HEP and NP and recently astrophysics & cosmology. Of course PLB is not the only journal that covers those fields, but it is unique in that it covers these fields in one journal, without distinguishing HEP and NP. It is a letter journal, so manuscripts are typically limited to 5-7 pages and are intended as a way to communicate results in a short and comprehensive manner. This is the beginning of an effort to try and get an estimate of the number of publications per year from the field of particle physics, starting in the 1960's. The reason for starting with PLB is simply that I am familiar with the journal.

*Method and Results*
Elsevier's PLB is a scientific publication that is available for a fee, except for publications in the last few years which are available for free as part of the Open Access model. Nevertheless, all volumes of PLB are available online and for each manuscript the title, authors and an abstract are available to anybody. This publicly available information was used to accumulate the numbers in here.
      Starting with volume 24, each volume could consist of several issues. In 1967 for example there were three volumes (24,25 & 26 partially) covering 27 issues. An issue was typically published every 2.5 weeks. This was mainly driven by the fact that issues only appeared in print at that time. Over the years the number of issues per volume decreased from initially 15 to a handful by 2013. Starting with volume 728 in January 2014 issue numbers were abandoned and typically a volume started covering one month worth of articles. By that time nearly all information was online.
      Since 1967 an article in PLB was identified by volume, issue, starting page and year. Page numbers started at zero for any new volume and continued increasing over all issues in a volume. So in principle the issue number is not needed. Starting with volume 728 in January 2014, articles were

identified by volume, starting page and year. Starting with volume 797 in 2019 page numbers were abandoned and an article is identified by volume, article ID and year.

PLB has a designation for each article. The possible designations are "Short communication" ( mostly used after 2000), "Research article" ( mostly used before 2000), "Discussion"( rarely used, but over all years), "Editorial"( news from publisher), "Review article ( mostly publication of the Particle Data Group) and 'Erratum" ( for corrections). Furthermore, each Volume/issue lists the Editorial Board at its beginning and that article has no designation. For this report only articles that are described by "Short Communication" ( newer), "Research Article"(older description) and "Discussion" are counted. So for example the publication of the Editorial Board and errata are ignored.

PLB started grouping articles into subject areas with volume 600 in 2004. Before that there were no groupings and all articles were simply listed. These subject areas are Astrophysics & Cosmology (astro), Experiments for experimental results (expe), Phenomenology (phen) and Theory (theor). It is up to the authors and editors to decide in which subject area an article is published. As far as I know there are no hard rules. There was an attempt to start subject areas (it seems) in 1988 in Volume 206, only in Issue 3, by splitting the publications in two subject areas : 1)Nuclei and 2)Particles & Fields.
In 2019 in Volume 793 there is a subject area "Short Communications", which contains all types of articles and of various length. It is simply assigned to "phen" for this report. At a few instances between 2005 and 2013, some publications were assigned to a subject area: " Comments", which are either comments on publications or comments on comments. There are nine of them, and they are ignored. In 2012, Volume 716, Issue 1, PLB published the 2 papers about the discovery of the Higgs boson at the CERN LHC in a separate subject area called: "Observation of a new particle in the search for the Standard Model Higgs boson". This was simply assigned to "expe'.

The data presented in here were obtained by scripts written which downloaded each web page (for a volume and some issues) from the PLB webpages and extracting the volume, year and issue number. For each article found the title, the subject area (if it existed), starting page or article ID and some other information were stored in simple text files, that could easily be analyzed and interpreted many times, ending up with a compete index of all article in PLB since its start. One can do many things with the data. For now, simply the number of publications are counted for each year and the results are given in Table 1.

| Year | PLB None | PLB Nucl | PLB Part | PLB Total | PLB Pages | Year | PLB Astro | PLB Expe | PLB Phen | PLB Theo | PLB None | PLB Total | PLB Pages |
|---|---|---|---|---|---|---|---|---|---|---|---|---|---|
| 1967 | 510 | | | 510 | 1578 | 1995 | | | | | 1571 | 1571 | 10717 |
| 1968 | 529 | | | 529 | 1652 | 1996 | | | | | 1653 | 1653 | 11241 |
| 1969 | 535 | | | 535 | 1687 | 1997 | | | | | 1546 | 1546 | 10888 |
| 1970 | 566 | | | 566 | 1911 | 1998 | | | | | 1766 | 1766 | 12941 |
| 1971 | 693 | | | 693 | 2432 | 1999 | | | | | 1436 | 1436 | 10484 |
| 1972 | 784 | | | 784 | 2969 | 2000 | | | | | 1382 | 1382 | 10383 |
| 1973 | 720 | | | 720 | 2645 | 2001 | | | | | 1335 | 1335 | 10097 |
| 1974 | 724 | | | 724 | 2701 | 2002 | | | | | 1155 | 1155 | 8910 |
| 1975 | 697 | | | 697 | 2678 | 2003 | | | | | 968 | 968 | 7590 |
| 1976 | 749 | | | 749 | 2815 | 2004 | 12 | 17 | 65 | 46 | 898 | 1038 | 8335 |
| 1977 | 834 | | | 834 | 3214 | 2005 | 124 | 152 | 426 | 253 | | 955 | 7638 |
| 1978 | 892 | | | 892 | 3573 | 2006 | 158 | 135 | 444 | 263 | | 1000 | 6122 |
| 1979 | 954 | | | 954 | 3962 | 2007 | 89 | 105 | 385 | 263 | | 842 | 5372 |
| 1980 | 954 | | | 954 | 4057 | 2008 | 134 | 100 | 404 | 291 | | 929 | 5323 |
| 1981 | 1097 | | | 1097 | 4810 | 2009 | 125 | 115 | 362 | 328 | | 930 | 5174 |
| 1982 | 1200 | | | 1200 | 5280 | 2010 | 133 | 94 | 293 | 249 | | 769 | 4489 |
| 1983 | 1473 | | | 1473 | 6724 | 2011 | 124 | 145 | 427 | 314 | | 1010 | 6232 |
| 1984 | 1602 | | | 1602 | 7561 | 2012 | 106 | 168 | 319 | 276 | | 869 | 6024 |
| 1985 | 1471 | | | 1471 | 7146 | 2013 | 76 | 140 | 282 | 281 | | 779 | 5356 |
| 1986 | 1460 | | | 1460 | 7298 | 2014 | 113 | 114 | 299 | 290 | | 816 | 5126 |
| 1987 | 1605 | | | 1605 | 8434 | 2015 | 88 | 133 | 295 | 311 | | 827 | 5486 |
| 1988 | 1578 | 5 | 30 | 1613 | 8493 | 2016 | 85 | 147 | 358 | 290 | | 880 | 6306 |
| 1989 | 1643 | | | 1643 | 8856 | 2017 | 88 | 141 | 337 | 311 | | 877 | 6241 |
| 1990 | 1686 | | | 1686 | 9746 | 2018 | 68 | 142 | 313 | 326 | | 849 | 6178 |
| 1991 | 1742 | | | 1742 | 10389 | 2019 | 87 | 126 | 267 | 327 | | 807 | 4616 |
| 1992 | 1721 | | | 1721 | 10649 | | | | | | | | |
| 1993 | 1617 | | | 1617 | 10224 | Sums | 1610 | 1974 | 5276 | 4419 | 45231 | 58545 | 340348 |
| 1994 | 1485 | | | 1485 | 9595 | | | | | | | | |

Table 1: Articles published in PLB by year. For years 1967 to 2003, publications are not grouped into subject areas and total number of articles ("PLB total") are the same as articles without a subject area ("PLB none"). The "PLB pages" column is the number of pages published in a given year.

Some more notes on Table 1:
- The total number of articles/publications in PLB over this time period is 58545 ( with selections explained above).
- The subject areas started towards the end of 2004 and that is why there are entries for the four subject areas and for articles without a subject area ("PLB none") for that year.

The figures below simply display the above numbers graphically.

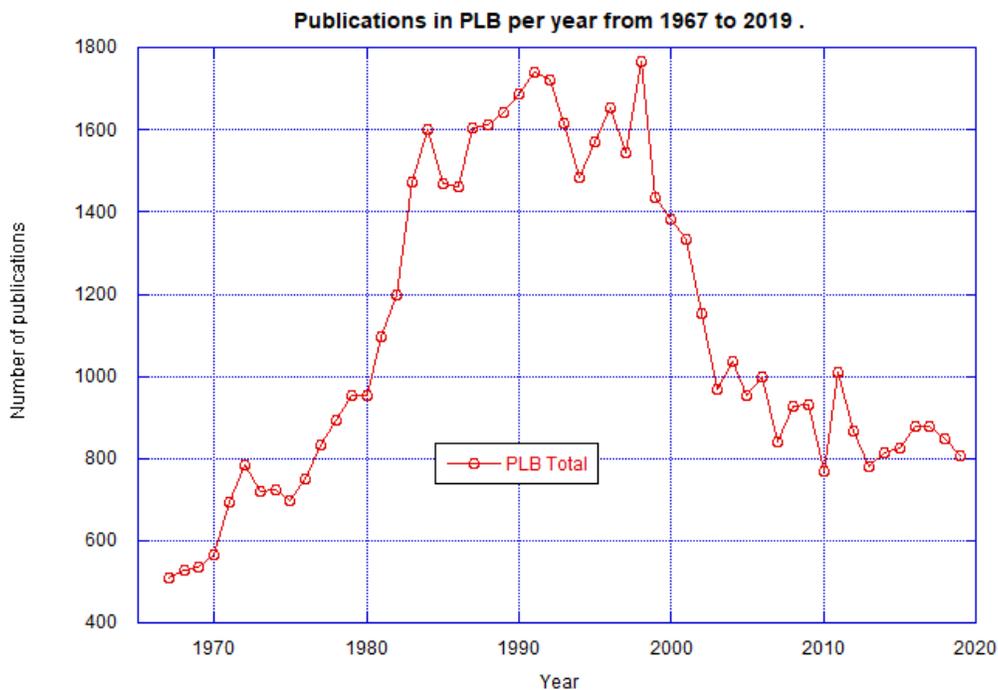

Figure 1: number of articles/publication per year for PLB.

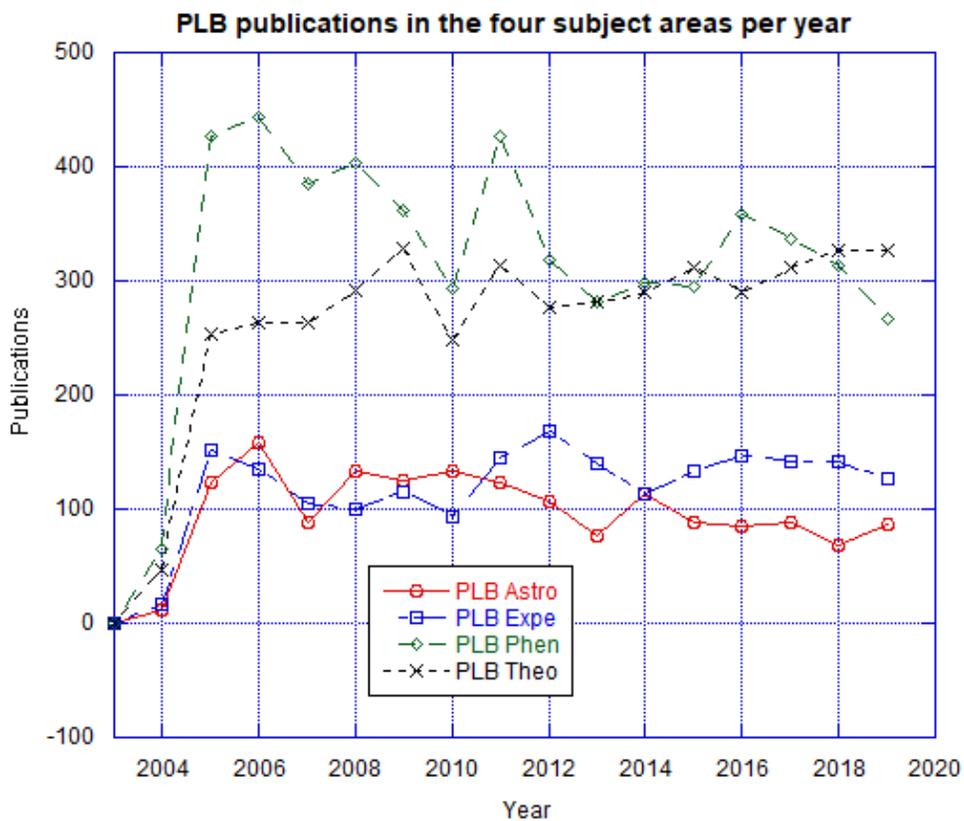

Figure 2: number of articles/publication per year for PLB in each of the subject areas, since they were introduced.

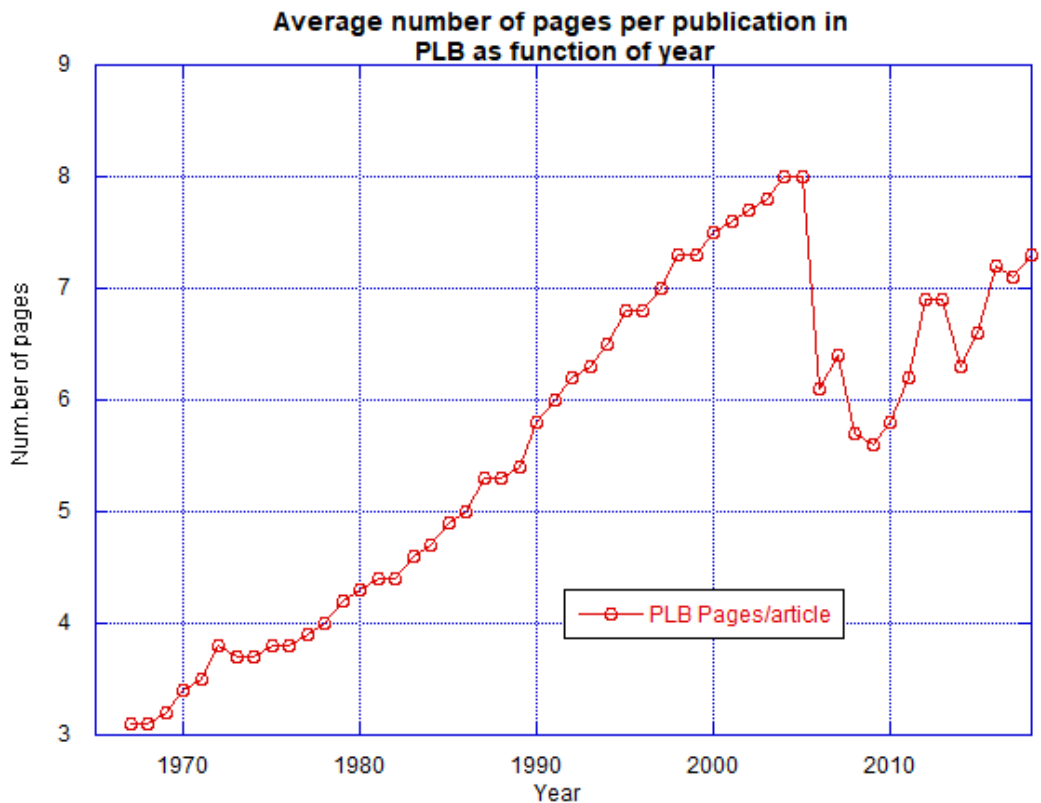

Figure 3: The number of pages per article/publication as a function of year. The last year shown is 2018, because in 2019 PLB changed to article ID and the number of pages for each article are not readily available.

*Conclusions*

Some conclusions from these graphs and numbers:
1. Obviously the publications per year grew steadily from about 1967 to 1982 or so, when they flatten (Figure 1).
2. They hold steady from ~1983 to ~1999 and they drop off to a new, much lower level. This may be due to the appearance of new letter journals which started accepting manuscripts, that normally would have gone to PLB (competition?) (Figure 1). This will be addressed in future reports covering other journals.
3. Ever since about 2003 the publications/articles per year seems steady, although there is a slight decrease with time (Figure 1).
4. The graph showing the publications per subject area, since 2004, basically is flat with time for each area. There is at least one spike in that graph and one might speculate whether that is related to some development in the field (Figure 2). Obviously publications in Phenomenology (phen) and Theory (theor) dominate.
5. Figure 3 shows the average number of pages per article for each year. There is a steady increase from 1967 from ~ 3 pages/article to 8 in 2005. The field has become more "wordy". Something happened around 2005 and the numbers drops to 6. It then remains about steady. It is not clear what causes these changes.

Version 1.0; July 2020
Publication: hjmw-02*Future steps*

      The next steps will be to analyze other journals which have published results from research in particle physics to get a more complete picture of the total number of articles resulting from the field of particle physics. In order to do this, it will be necessary to separate articles from HEP and NP. As a first step we will try and use the repositories in the arXiv [3] to get a first handle on this. Of course the arXiv started around 1992 and so the initial information will only cover the last three decades, but it will be a start.

*Acknowledgements & disclaimer:*

The support from COFI, allowing me to use the facility in San Juan, PR for a while enabling the download of webpages and initial development of the scripts used. This work was not supported by external funding and simply has arisen from a personal interest.

The information in this report may be available via other means and there is no claim of it being unique. This information may exist in some other form, but I was not able to easily find it. I simply pursued an interest in getting a historical perspective of PLB. Data used for this are available to anybody upon request.

*References*

[1] PLB, "All issues of Physics Letters B," *https://www.sciencedirect.com/journal/physics-letters-b/issues,* All years.

[2] Wikipedia, "History of Physics Letters," *https://en.wikipedia.org/wiki/Physics_Letters,* -.

[3] arXiv, "ArXiv and its repositories," *https://arxiv.org/,* All years.6